\documentclass[12pt,fleqn]{iopart}

\usepackage{amsmath}
\usepackage{bm}
\usepackage[dvips]{graphicx}
\usepackage{cite}

\newcommand{\fourier}[1]{\mathcal{F} \biggl[#1 \biggl]}
\newcommand{\ifourier}[1]{\mathcal{F}^{-1} \biggl[#1 \biggl]}
\newcommand{\sfourier}[1]{\mathcal{F} \bigl[#1 \bigr]}

\newcommand{\dop}[3]{{{\rm D}^{#1}_{#2} [#3]}} 

\begin{document}

\title{Calculation reduction method for color computer-generated hologram using color space conversion}

\author{Tomoyoshi Shimobaba$^1$, Takashi Kakue$^1$, Minoru Oikawa$^1$, \\
Naoki Takada$^2$, Naohisa Okada$^1$, Yutaka Endo$^1$, 
Ryuji Hirayama$^1$, \\
and Tomoyoshi Ito$^1$}

\address{
$^1$Graduate School of Engineering, Chiba University, 1-33 Yayoi-cho, Inage-ku, Chiba 263-8522, Japan\\
$^2$Faculty of Science, Kochi University, 2--5--1, Akebono--cho, Kochi 780--8520, Japan\\
$^*$Corresponding author: shimobaba@faculty.chiba-u.jp
}
\begin{abstract}
We report a calculation reduction method for color computer-generated holograms (CGHs) using color space conversion.
Color CGHs are generally calculated on RGB space.
In this paper, we calculate color CGHs in other color spaces: for example, YCbCr color space.
In YCbCr color space, a RGB image is converted to the luminance component (Y), blue-difference chroma (Cb) and red-difference chroma (Cr) components.
In terms of the human eye, although the negligible difference of the luminance component is well-recognized, the difference of the other components 
is not.
In this method,  the luminance component is normal sampled and the chroma components are down-sampled.
The down-sampling allows us to accelerate the calculation of the color CGHs.
We compute diffraction calculations from the components, and then we convert the diffracted results in YCbCr color space to RGB color space. 
\end{abstract}


\maketitle

\section{Introduction}
In recent years, three dimensional (3D) displays using computer-generated holograms (CGHs) have been well studied, since CGHs can faithfully reconstruct the light waves of 3D images \cite{poon}. 
Such 3D display is referred as to ``Electroholography''.
Electroholography can reconstruct 3D images by displaying CGHs generated by diffraction calculations on a spatial light modulator (SLM) such as amplitude or phase-modulated LCD panels.
In electroholography, several important issues must be considered: calculation cost for CGH, narrow viewing angle, small reconstructed images and color reconstruction.

As color reconstruction is an important issues, many color reconstruction methods have been studied. 
For instance, color reconstruction methods using three SLMs to display three CGHs corresponding to the red, green and blue components of a 3D image have been proposed \cite{Sato, Onual, Onual1, Yoshikawa, Nakayama}. 
Color reconstruction methods using a single SLM have also been proposed:   space-division \cite{Ito1,Ito2} and depth-division \cite{Makowski1,Makowski2} methods. 
Another single SLM method is the time-division method, which temporally switches RGB lights by synchronizing signals \cite{Shimo1,Shimo2,Shimo3,Shimo4}. 
Regardless of the color reconstruction methods as mentioned above, we need to compute three diffraction calculations corresponding to the RGB components of a color image.

We developed a calculation reduction method for color CGHs using color space conversion.
Color CGHs have been calculated on RGB space.
In this paper, we calculate color CGHs in other color spaces: for example,  YCbCr color space, which is widely used in digital image processing like JPEG and MPEG formats.
In YCbCr color space, a RGB image is converted to the luminance component (Y), blue-difference chroma (Cb) and red-difference chroma (Cr) components.
In terms of the human eye, although the negligible difference of the luminance component is well-recognized, the difference of the other components 
is not.
In this method,  the luminance component is normal sampled and the chroma components are down-sampled.
The down-sampling allows us to accelerate the calculation of the color CGHs.
We calculate diffraction from the components, and then convert the diffracted result in YCbCr color space to RGB color space. 

In Section 2, we explain the calculation reduction method for color CGH using color space conversion.
In Section 3, we verify the proposed method on computer simulation.
Section 4 concludes this work.

\section{Calculation reduction method for color computer-generated hologram using color space conversion}

For simple notation we introduce diffraction operator $\dop{z}{\lambda}{u(\bm x)}$, where $z$ is the propagation distance, $\lambda$ is the wavelength, $u(\bm x)$ is a source plane and  $\bm x = (x,y)$ is the position vector.
For example, the angular spectrum method \cite{goodman}, which is a major diffraction calculation method, is expressed by the diffraction operator as follows: 
\begin{eqnarray}
u_2(\bm x_2) &=& \dop{z}{\lambda}{u_1(\bm x_1)}  = \int \fourier{u_1(\bm x_1)}  \exp \biggl( 2 \pi i z \sqrt{\frac{1}{\lambda^2} - |\bf f|^2} ~ \biggr)   \exp(2\pi i {\bf f x_2}) d {\bm f} \nonumber \\
&=& \ifourier{ \fourier{u_1(\bm x_1)}  \exp \biggl( 2 \pi i z \sqrt{\frac{1}{\lambda^2} - |\bf f|^2} ~ \biggr) },
\label{eqn:fre}
\end{eqnarray}
where  $\bm f=(f_x,f_y)$ is the position vector in Fourier domain, $u_1(\bm x_1)$ is a source plane, $u_2(\bm x_2)$ is a destination plane, $\sfourier{\cdot}$ is Fourier transform, and $z$ is the propagation distance between the source and destination planes.

If we calculate color CGHs in RGB color space, we need to calculate three diffracted fields $u_R(\bm x_2), u_G(\bm x_2)$ and $u_B(\bm x_2)$ from the corresponding components of the RGB images, $R(\bm x_1), G(\bm x_1)$ and $B(\bm x_1)$ as follows:
\begin{eqnarray}
u_R(\bm x_2)=\dop{z}{\lambda_R}{R(\bm x_1)} \label{eqn:rgb_cgh1},\\
u_G(\bm x_2)=\dop{z}{\lambda_G}{G(\bm x_1)} \label{eqn:rgb_cgh2},\\
u_B(\bm x_2)=\dop{z}{\lambda_B}{B(\bm x_1)} \label{eqn:rgb_cgh3},
\end{eqnarray}
where $\lambda_R, \lambda_G$ and $\lambda_B$ are the wavelengths of RGB lights, respectively. 
In this paper, we call this the ``direct color CGH calculation''.
If there are spatial light modulators that are capable of displaying complex amplitudes \cite{complex_slm}, we can reconstruct a clear RGB image without direct and conjugated lights from Eqs.(\ref{eqn:rgb_cgh1})-(\ref{eqn:rgb_cgh3}) by the three SLM method \cite{Sato, Onual, Onual1, Yoshikawa, Nakayama}, time-division method \cite{Shimo1,Shimo2,Shimo3,Shimo4} and space-division method \cite{Ito1,Ito2} and so forth.
When using the amplitude or phase-modulated SLMs, we need to convert the complex amplitudes of Eqs.(\ref{eqn:rgb_cgh1})-(\ref{eqn:rgb_cgh3}) to the amplitude CGHs by taking the real part or phase-only CGH by taking the argument.

From here, let us consider the calculation of color CGHs in the YCbCr color space. 
Converting a RGB image to a YCbCr image is as follows,
\begin{equation}
\begin{pmatrix}
Y({\bm x}) \\
C_b({\bm x}) \\
C_r({\bm x})
\end{pmatrix}=
\begin{pmatrix}
a_{11} & a_{12} & a_{13} \\
a_{21} & a_{22} & a_{23} \\
a_{31} & a_{32} & a_{33} \\
\end{pmatrix}
\begin{pmatrix}
R({\bm x}) \\
G({\bm x}) \\
B({\bm x})
\end{pmatrix}=
\begin{pmatrix}
0.2989 & 0.5866 & 0.1145 \\
-0.1687 & -0.3312 & 0.5 \\
0.5 & -0.4183 & -0.0816 \\
\end{pmatrix}
\begin{pmatrix}
R({\bm x}) \\
G({\bm x}) \\
B({\bm x})
\end{pmatrix},
\label{eqn:r_y}
\end{equation}
where $Y(\bm x)$, $C_b(\bm x)$ and $C_r(\bm x)$ are the luminance, blue-difference chroma and red-difference chroma, respectively. 
We convert the YCbCr image to the RGB image by,
\begin{equation}
\begin{pmatrix}
R({\bm x}) \\
G({\bm x}) \\
B({\bm x})
\end{pmatrix}=
\begin{pmatrix}
b_{11} & b_{12} & b_{13} \\
b_{21} & b_{22} & b_{23} \\
b_{31} & b_{32} & b_{33} \\
\end{pmatrix}
\begin{pmatrix}
Y({\bm x}) \\
C_b({\bm x}) \\
C_r({\bm x})
\end{pmatrix}=
\begin{pmatrix}
1 & 0 & 1.4022 \\
1 & -0.3456 & -0.7145 \\
1 & 1.7710 & 0 \\
\end{pmatrix}
\begin{pmatrix}
Y({\bm x}) \\
C_b({\bm x}) \\
C_r({\bm x})
\end{pmatrix}.
\label{eqn:y_r}
\end{equation}

From Eq.(\ref{eqn:y_r}), the diffracted fields of RGB images passing through the YCbCr color space are expressed as,
\begin{eqnarray}
\dop{z}{\lambda_R}{R(\bm x_1)}=b_{11} \dop{z}{\lambda_R}{Y(\bm x_1)} + b_{12} \dop{z}{\lambda_R}{C_b(\bm x_1)} + b_{13} \dop{z}{\lambda_R}{C_r(\bm x_1)}  \label{eqn:diff_y_r1}, \\ 
\dop{z}{\lambda_G}{G(\bm x_1)}=b_{21} \dop{z}{\lambda_G}{Y(\bm x_1)} + b_{22} \dop{z}{\lambda_G}{C_b(\bm x_1)} + b_{23} \dop{z}{\lambda_G}{C_r(\bm x_1)}  \label{eqn:diff_y_r2}, \\ 
\dop{z}{\lambda_B}{R(\bm x_1)}=b_{31} \dop{z}{\lambda_B}{Y(\bm x_1)} + b_{32} \dop{z}{\lambda_B}{C_b(\bm x_1)} + b_{33} \dop{z}{\lambda_B}{C_r(\bm x_1)} \label{eqn:diff_y_r3}.
\end{eqnarray}

Even though the direct color CGH calculation of Eqs.(\ref{eqn:rgb_cgh1})-(\ref{eqn:rgb_cgh3}) requires only three diffraction calculations, we need to calculate nine diffraction calculations in Eqs. (\ref{eqn:diff_y_r1})-(\ref{eqn:diff_y_r3}).
To improve this problem we use the following important relation of the diffraction operator:
\begin{equation}
\dop{z}{\lambda_1}{u(\bm x)} = \dop{z\lambda_2/\lambda_1}{\lambda_2}{u(\bm x)} .
\label{eqn:ope}
\end{equation}

Applying the relation of $\dop{z \lambda_G/\lambda_R}{\lambda_G}{u(\bm x)} = \dop{z}{\lambda_R}{u(\bm x)}$ to Eq.(\ref{eqn:diff_y_r2}), the following equation is derived,
\begin{eqnarray}
\dop{z\lambda_G/\lambda_R}{\lambda_G}{G(\bm x_1)}&=&
b_{21} \dop{z \lambda_G/\lambda_R}{\lambda_G}{Y(\bm x_1)} + 
b_{22} \dop{z \lambda_G/\lambda_R}{\lambda_G}{C_b(\bm x_1)} + 
b_{23} \dop{z \lambda_G/\lambda_R}{\lambda_G}{C_r(\bm x_1)}, \nonumber \\
&=& 
b_{21} \dop{z}{\lambda_R}{Y(\bm x_1)} + 
b_{22} \dop{z}{\lambda_R}{C_b(\bm x_1)} + 
b_{23} \dop{z}{\lambda_R}{C_r(\bm x_1)}.
\label{eqn:diff_y_4}
\end{eqnarray}

Likewise, applying the relation of $\dop{z \lambda_B/\lambda_R}{\lambda_B}{u(\bm x)} = \dop{z}{\lambda_R}{u(\bm x)}$ to Eq.(\ref{eqn:diff_y_r3}), the following equation is derived,
\begin{eqnarray}
\dop{z\lambda_B/\lambda_R}{\lambda_B}{B(\bm x_1)}&=&
b_{31} \dop{z \lambda_B/\lambda_R}{\lambda_B}{Y(\bm x_1)} + 
b_{32} \dop{z \lambda_B/\lambda_R}{\lambda_B}{C_b(\bm x_1)} + 
b_{33} \dop{z \lambda_B/\lambda_R}{\lambda_B}{C_r(\bm x_1)}, \nonumber \\
&=& 
b_{31} \dop{z}{\lambda_R}{Y(\bm x_1)} + 
b_{32} \dop{z}{\lambda_R}{C_b(\bm x_1)} + 
b_{33} \dop{z}{\lambda_R}{C_r(\bm x_1)}.
\label{eqn:diff_y_5}
\end{eqnarray}

Finally, we can obtain a color CGH passing through the YCbCR color space in matrix notation as follows:
\begin{equation}
\begin{pmatrix}
\dop{z}{\lambda_R}{R(\bm x_1)} \\
\dop{z\lambda_G/\lambda_R}{\lambda_G}{G(\bm x_1)} \\
\dop{z\lambda_B/\lambda_R}{\lambda_B}{B(\bm x_1)}
\end{pmatrix}=
\begin{pmatrix}
b_{11} & b_{12} & b_{13} \\
b_{21} & b_{22} & b_{23} \\
b_{31} & b_{32} & b_{33} \\
\end{pmatrix}
\begin{pmatrix}
\dop{z}{\lambda_R}{Y(\bm x_1)} \\
\dop{z}{\lambda_R}{C_b({\bm x_1})} \\
\dop{z}{\lambda_R}{C_r({\bm x_1})}
\end{pmatrix}.
\label{eqn:diff_y_r_mat}
\end{equation}

As you can see, this conversation includes only three diffraction operators.
Unfortunately, the propagation distance of the diffracted results of green and blue components change from $z$ to $z \lambda_G/\lambda_R$ and  $z \lambda_B/\lambda_R$.
When placing CGHs corresponding to each component at same the location, the reconstructed RGB images from the CGHs are reconstructed out of position.
We can compensate for the out of location by placing the green and blue CGHs at $z \lambda_R/\lambda_G$ and  $z \lambda_R/\lambda_B$, respectively.

Since Eq.(\ref{eqn:diff_y_r_mat}) has three diffraction operators, the calculation cost is the same as the direct calculation of the direct color CGH calculation by Eqs.(\ref{eqn:rgb_cgh1})-(\ref{eqn:rgb_cgh3}).
In terms of the human eye, although the negligible difference of the luminance component Y is well-recognized, the difference of the other components Cb and Cr is not.
Using this property, the luminance component Y is normal sampled and the chroma components are down-sampled.
The down-sampling allows us to accelerate the calculation of the color CGHs.
Note that we need to change the sampling pitches that are required for the diffraction operator of down-sampled $C_b(\bm x_1)$ and $C_b(\bm x_1)$ because the areas of $C_b(\bm x_1)$ and $C_b(\bm x_1)$ are smaller than   $Y(\bm x_1)$.
For example, when down-sampling 1/4 $C_b(\bm x_1)$ and $C_b(\bm x_1)$, we change the sampling pitches of them to 4$p$ where $p$ is the sampling pitch of $Y(\bm x_1)$.

Let us summarize the color CGH calculation passing through the YCbCr color space.
First, we convert a RGB image to a YCbCr image using Eq.(\ref{eqn:r_y}).
Second, we down-sample the Cb and Cr components.
Third, we compute diffraction calculations from the components, then, we up-sample $\dop{z}{\lambda_R}{C_b(\bm x_1)} $ and $\dop{z}{\lambda_R}{C_r(\bm x_1)}$.
Next, we calculate the RGB CGHs using Eq.(\ref{eqn:diff_y_r_mat}).
Last, we reconstruct the color images by placing the green and blue CGHs at $z \lambda_R/\lambda_G$ and  $z \lambda_R/\lambda_B$, respectively.

\section{Results}

\begin{figure}[htb]
\centerline{
\includegraphics[width=15cm]{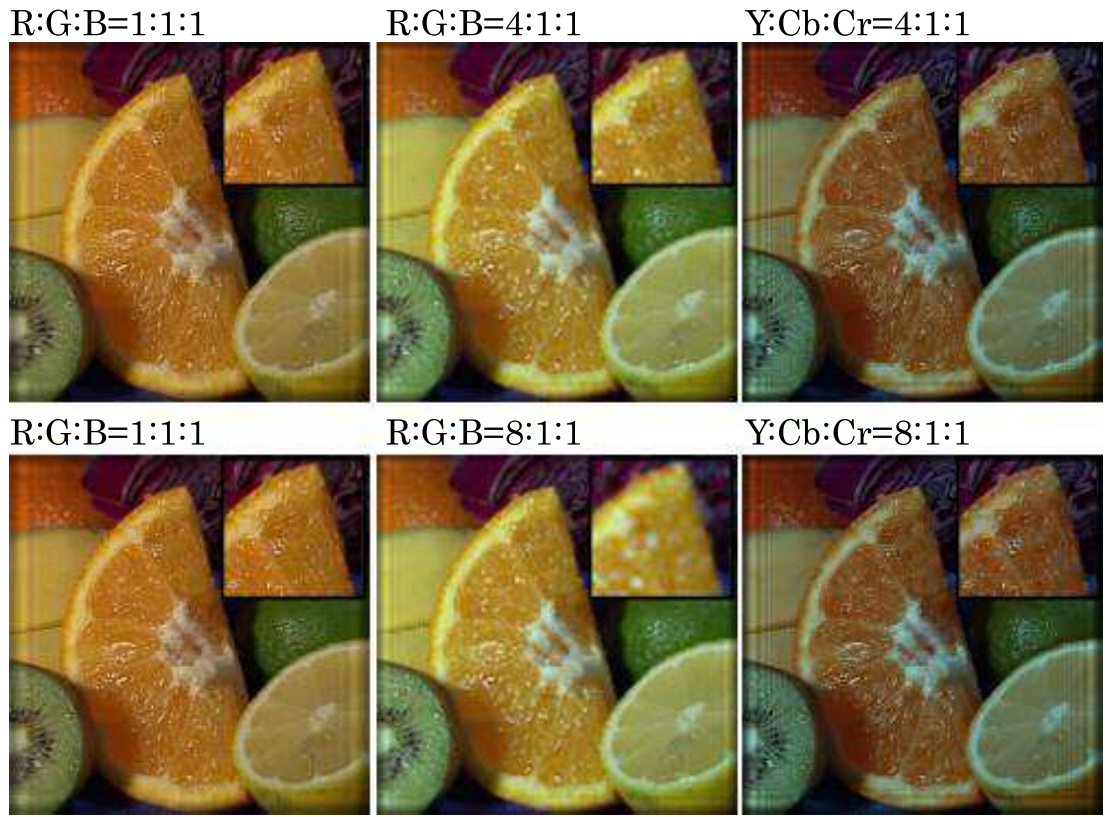}}
\caption{Reconstructed images of ``Fruits'' from CGHs, which are calculated by direct color CGH calculation (Eqs.(\ref{eqn:rgb_cgh1})-(\ref{eqn:rgb_cgh3})) and the proposed method (Eq.(\ref{eqn:diff_y_r_mat})), respectively}
\label{fig:img1}
\end{figure}

We verify the proposed method on computer simulation.
We use three RGB color images.
For the calculation condition, the propagation distance $z=5$cm, the sampling pitch is 10$\mu m$, the wavelengths of RGB lights are 633 nm, 532 nm and 450 nm, respectively.
The resolution of all of the images is $512 \times 512$ pixels.
We assume that we use a complex amplitude SLM \cite{complex_slm}.
We use the nearest-neighbor interpolation for the down-sampling and up-sampling in the color space conversion.
The nearest-neighbor interpolation is simple interpolation, so that the interpolation is faster than other interpolation methods such as linear and cubic interpolations.

Figure \ref{fig:img1} shows the reconstructed images of ''Fruits'' from CGHs, which are calculated by direct color CGH calculation (Eqs.(\ref{eqn:rgb_cgh1})-(\ref{eqn:rgb_cgh3})) and the proposed method (Eq.(\ref{eqn:diff_y_r_mat})), respectively.

The notations of ``R:G:B=1:1:1'', ``R:G:B=4:1:1'' mean a color image is not down-sampled and is down-sampled by 1/4 of the green and blue components, respectively. 
Likewise,  The notation of ``Y:Cb:Cr=4:1:1'' mean that a color image is down-sampled by 1/4 of Cb and Cr components, respectively.  
The insects show magnified images of a part of the reconstructed images.

The reconstructed images of  ``R:G:B=4:1:1'' and ``R:G:B=8:1:1'' are blurred because the green and blue components affect the luminance of the image. 
In contrast, the reconstructed images of  ``Y:Cb:Cr=4:1:1'' and `` Y:Cb:Cr=8:1:1'' maintain the sharpness of the texture because Cb and Cr components do not seriously affect the luminance
However, the brightness is little decreased, compared with other reconstructed images. 

\begin{figure}[htb]
\centerline{
\includegraphics[width=15cm]{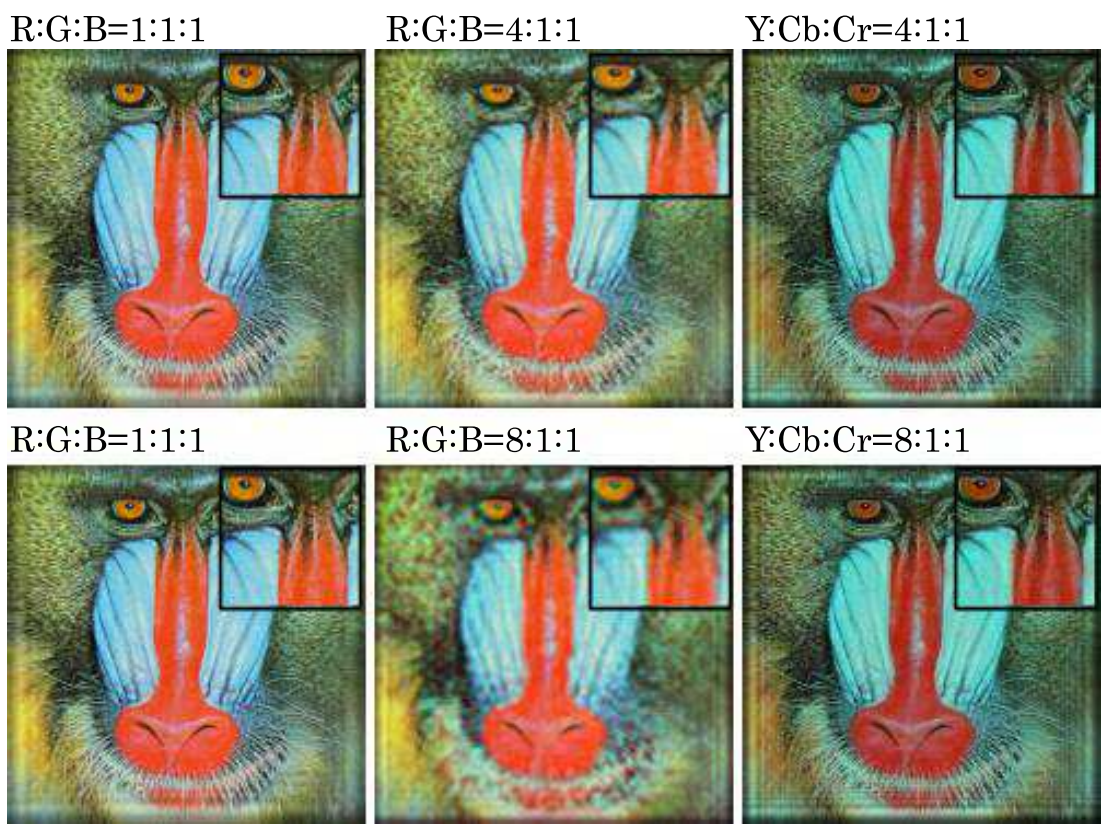}}
\caption{Reconstructed images of ``Mandrill'' from CGHs, which are calculated by direct color CGH calculation (Eqs.(\ref{eqn:rgb_cgh1})-(\ref{eqn:rgb_cgh3})) and the proposed method (Eq.(\ref{eqn:diff_y_r_mat})), respectively}
\label{fig:img2}
\end{figure}

\begin{figure}[htb]
\centerline{
\includegraphics[width=15cm]{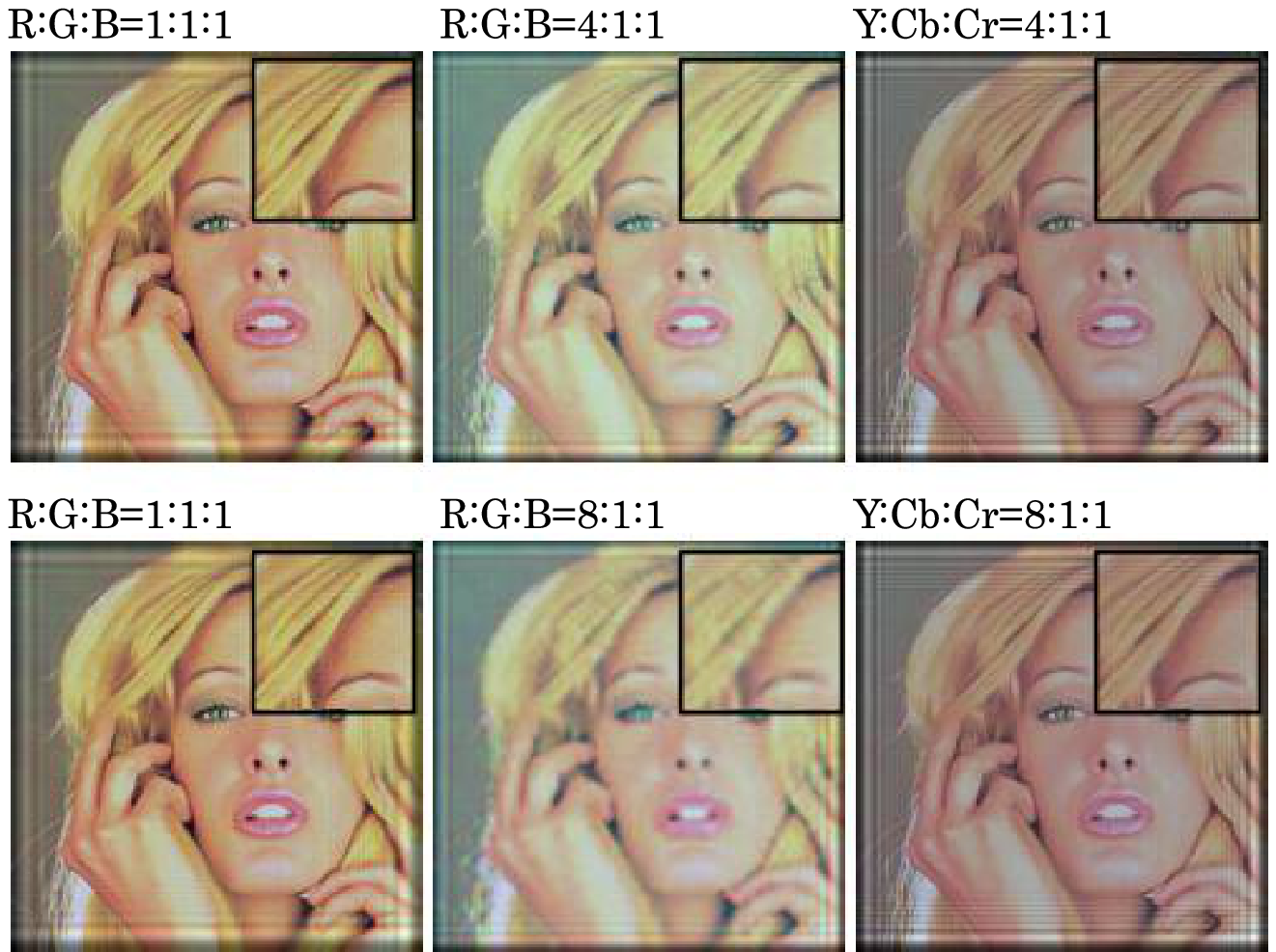}}
\caption{Reconstructed images of ``Tifanny'' from CGHs, which are calculated by direct color CGH calculation (Eqs.(\ref{eqn:rgb_cgh1})-(\ref{eqn:rgb_cgh3})) and the proposed method (Eq.(\ref{eqn:diff_y_r_mat})), respectively}
\label{fig:img3}
\end{figure}

Figures \ref{fig:img2} and \ref{fig:img3} show the reconstructed images of ''Mandrill'' and ``Tifanny'' from CGHs, which are calculated by direct color CGH calculation (Eqs.(\ref{eqn:rgb_cgh1})-(\ref{eqn:rgb_cgh3})) and the proposed method (Eq.(\ref{eqn:diff_y_r_mat})), respectively.
The reconstructed images of  ``R:G:B=4:1:1'' and `` R:G:B=8:1:1'' are blurred.
In contrast, the reconstructed images of  ``Y:Cb:Cr=4:1:1'' and `` Y:Cb:Cr=8:1:1'' maintain the sharpness of the texture.

The calculation times of ``R:G:B=1:1:1'', ``R:G:B=4:1:1'' and ``Y:Cb:Cr=4:1:1'' are about 660 ms, 270 ms and 396 ms, respectively.
The calculation times of ``R:G:B=8:1:1'' and ``Y:Cb:Cr=8:1:1'' are about 260 ms and 400 ms, respectively.
Therefore, the proposed method can maintain the sharpness of the reconstructed image and accelerate the calculation speed.

\section{Conclusion}
We proposed a calculation reduction method for color CGHs using color space conversion.
The proposed method succeeded in maintaining the sharpness of the reconstructed image and accelerating the calculation speed.
The proposed method will be acceptable not only for YCbCr color space but also other color spaces such as YIQ, YUV and so forth.   
In addition, the proposed method will be useful for color digital holography to accelerate the numerical reconstruction. 
The proposed method applies only to a pure amplitude image or weak phase image, so in future study we will attempt to develop a calculation reduction method for color CGHs with full complex amplitude.

\section*{Acknowledgement}
This work is supported by Japan Society for the Promotion of Science (JSPS) KAKENHI (Grant-in-Aid for Scientific Research (C) 25330125) 2013, and KAKENHI (Grant-in-Aid for Scientific Research (A) 25240015) 2013.

\end{document}